\documentclass[conference]{IEEEtran}

\usepackage{epsfig}
\usepackage{multirow}
\usepackage{microtype} 
\usepackage{booktabs}  
\usepackage{url}
\usepackage{paralist}
\usepackage{subfig}
\usepackage{todonotes}
\usepackage{amsmath}
\usepackage{mathtools}
\usepackage{algpseudocode}
\usepackage{algorithm}
\usepackage{algorithmicx}
\usepackage{url}
\usepackage{paralist}  

\newcommand{\overbar}[1]{\mkern 1.5mu\overline{\mkern-1.5mu#1\mkern-1.5mu}\mkern 1.5mu}

\begin{document}

\date{}

\title{Threshold-based Obfuscated Keys with Quantifiable Security against Invasive Readout}

\author{\IEEEauthorblockN{Shahrzad Keshavarz and Daniel Holcomb}
	\IEEEauthorblockA{\textit{Department of Electrical and Computer Engineering} \\
		\textit{University of Massachusetts Amherst}\\
		{\{skeshavarz,holcomb\}@engin.umass.edu}}
	}

\maketitle

\begin{abstract}
Advances in reverse engineering make it challenging to deploy any on-chip information in a way that is hidden from a determined attacker. A variety of techniques have been proposed for design obfuscation including look-alike cells in which functionality is determined by hard to observe mechanisms including dummy vias or transistor threshold voltages. Threshold-based obfuscation is especially promising because threshold voltages cannot be observed optically and require more sophisticated measurements by the attacker. In this work, we demonstrate the effectiveness of a methodology that applies threshold-defined behavior to memory cells, in combination with error correcting codes to achieve a high degree of protection against invasive reverse engineering. The combination of error correction and small threshold manipulations is significant because it makes the attacker's job harder without compromising the reliability of the obfuscated key.  We present analysis to quantify key reliability of our approach, and its resistance to reverse engineering attacks that seek to extract the key through imperfect measurement of transistor threshold voltages. The security analysis and cost metrics we provide allow designers to make a quantifiable tradeoff between cost and security. We find that the combination of small threshold offsets and stronger error correcting codes are advantageous when security is the primary objective.

\end{abstract}

\section{Introduction}
\label{sec:introduction}
The goal of hardware obfuscation is to hide certain information about a design, making it harder for a reverse engineer to extract secret data or information about the function that is implemented. A designer may seek different objectives through hardware obfuscation. She might try to hide the structure of a circuit while preserving its functionality, or simply try to store a secret key in hardware. However as more sophisticated and advanced methods are proposed for design obfuscation, the reverse engineering procedures such as delayering and schematic reconstruction are becoming more capable, automated, and available~\cite{torrance-11, vijayakumar2016physical}. Designers wishing to keep secrets on chips have to find new and better approaches for stopping reverse engineering, leading to a continued arms race between designers and attackers. 

\subsection{Related Work}

In recent years, several approaches have been proposed for hardware obfuscation of logic. One notable approach is the use of look-alike cells with dummy contacts \cite{Rajendran_2013, chow2007integrated} that are intended to be hard for an attacker to read out invasively after delayering a chip. A variation of the same idea is the use of transistor thresholds or doping to change the function of cells as introduced by Becker et al.~\cite{becker2013stealthy}. The modification of threshold voltages can make transistors permanently on/off or can adjust their characteristics in a way that will change the function of a logic gate~\cite{NirmalaThreshold, Erbagci2016threshold, collantes2016threshold}. Design automation for logic obfuscation can be used to deploy the aforementioned modified transistors as basic library cells \cite{keshavarz2017design}. Note that all obfuscations techniques of this type induce the same logic function on all chip instances; this provides an attractive high-value target for a determined attacker. Although threshold voltages cannot be learned from optical analysis during reverse engineering, techniques do exist which can measure thresholds. Sugawara et al.~\cite{sugawara2015} show that it is possible to invasively read out information about transistor doping of the type utilized by Becker~\cite{becker2013stealthy}.

Similar to obfuscated logic that can be attacked invasively, data stored in flash memory or antifuses can be attacked and read out invasively when a chip is not powered. This is especially troublesome in the case of high-value master keys that must be stored on many instances of a chip. To mitigate the threat of invasive readout, Valamehr et al. present a scheme to protect data from invasive attack~\cite{valamehr2012inspection}; in their work, an attacker is required to read out a large number of cells correctly in order to extract the key, but the information read by the attacker is still digital information. By contrast, the approach we will present uses distributed analog secrets to prevent attack, in order to offer even stronger defense by not having any sensitive digital artifact that can be attacked at rest. 

\subsection{Proposed Approach}

An important idea in Cryptography is Kerckhoffs' principle -- that security should rely only on the key, and it should not matter whether an adversary knows the algorithm. In other words, a system should remain secure even if everything except the key is public knowledge. A common restatement of this principle is to avoid ``security through obscurity.'' In this work, we try to adhere to Kerckhoffs' principle by obfuscating only a key and allowing an attacker to know everything about the design except for the characteristics of certain transistors that determine the key. 
The advantages of doing so will be the following:
\begin{enumerate}
	\item Modularity: Involving an arbitrary logic circuit in obfuscation comes with extra difficulties for design and test. However, if the key is separate from the logic, the interactions between obfuscation and logic are minimal and well-defined. This allows for a detailed exploration of key obfuscation without regard to its impact on logic. 
	\item Generality: A key is perhaps the most general type of information to hide on a chip, and an obfuscated key can also be used in a straightforward way to obfuscate logic~\cite{yasin-transforming}. 
	\item Use of error correction: Most importantly, a separate obfuscated key storage allows for the use of error correction, which is not possible when logic is obfuscated directly. We will show that error correction is crucial for allowing the threshold differences to be small enough to fool the attacker without compromising reliability. 
\end{enumerate}

\begin{figure}
	\centering
	\includegraphics[width=0.8\columnwidth]{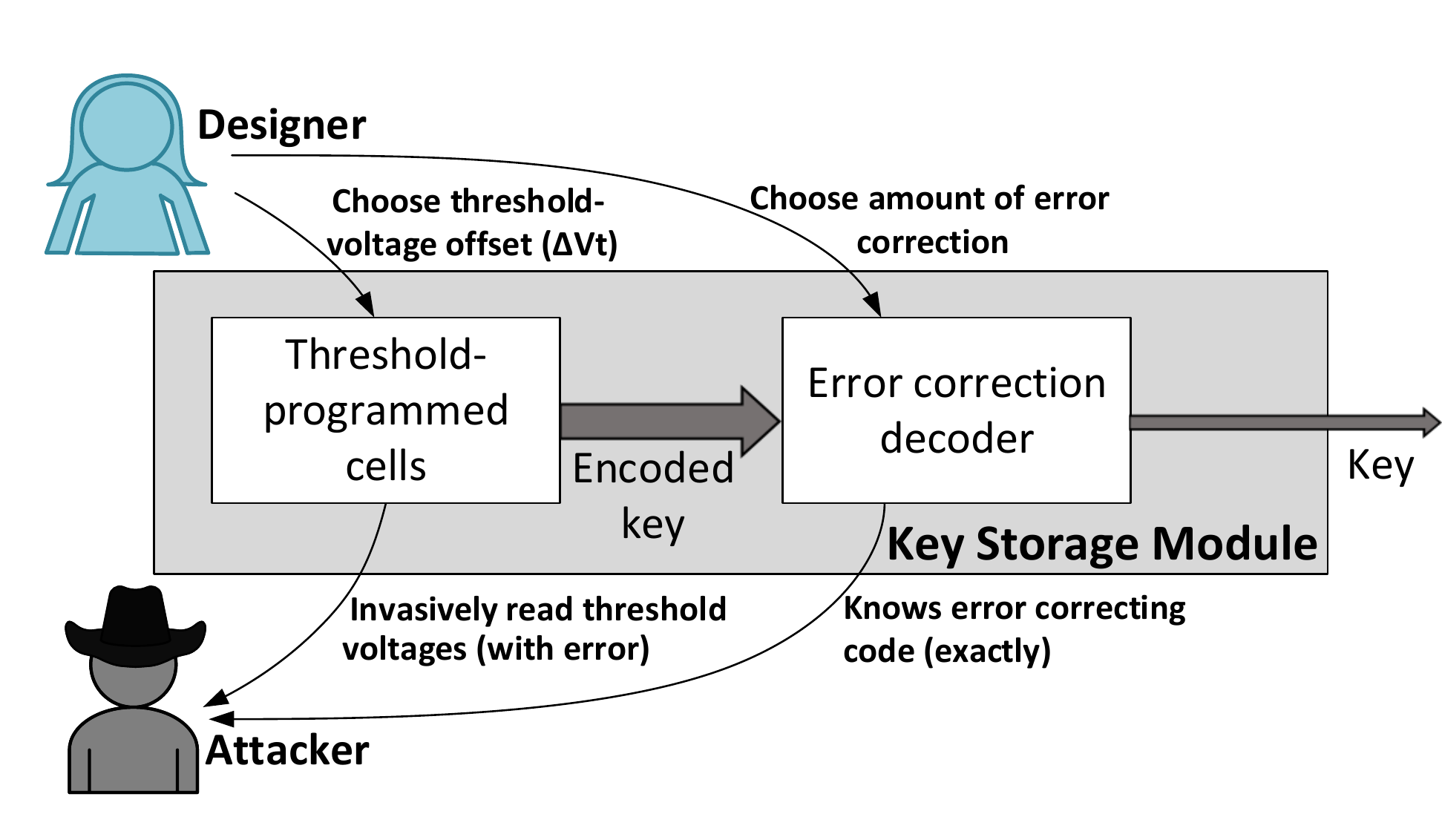}
	\caption{The overall design approach, both from the designer's and attacker's perspective}
	\label{fig:designerVSattacker}
\end{figure}

Figure \ref{fig:designerVSattacker} shows the overall key generation process from the designer and attacker perspective. The design consists of a block containing cells with modified threshold voltages, which is used to generate the encoded secret key when it is needed. The output of this block is given to an error correction block, which decodes the encoded secret key in a way that tolerates errors. 
The designer can choose the threshold offset of cells to store the encoded secret key, and can choose the parameters of the error correction codes to ensure key reliability. The attacker is allowed to know the error correction used, and knows the mechanism that the designer uses to configure the cells; the attacker can even invasively measure the threshold voltages of the cells, but does so imperfectly. If the attacker is able to get enough information about the cells, then he will be able to guess the encoded key accurately enough to produce the secret key by applying the known error correction scheme to it. 



The specific contributions of this work are as follows:
\begin{itemize}
\item We present the first approach that combines threshold-based obfuscation and error correction.
\item We show that this approach leads to quantifiable protection against invasive readout using a very conservative attacker model that only assumes some amount of imprecision when invasively measuring device threshold voltages.
\item We give a CAD flow for deploying the proposed approach in a way that can achieve various tradeoffs between reliability, security, and cost.
\end{itemize}

\section{Sketch of Approach}
\label{sec:sketch}

\begin{figure}
	\centering
	\includegraphics[width=0.95\columnwidth]{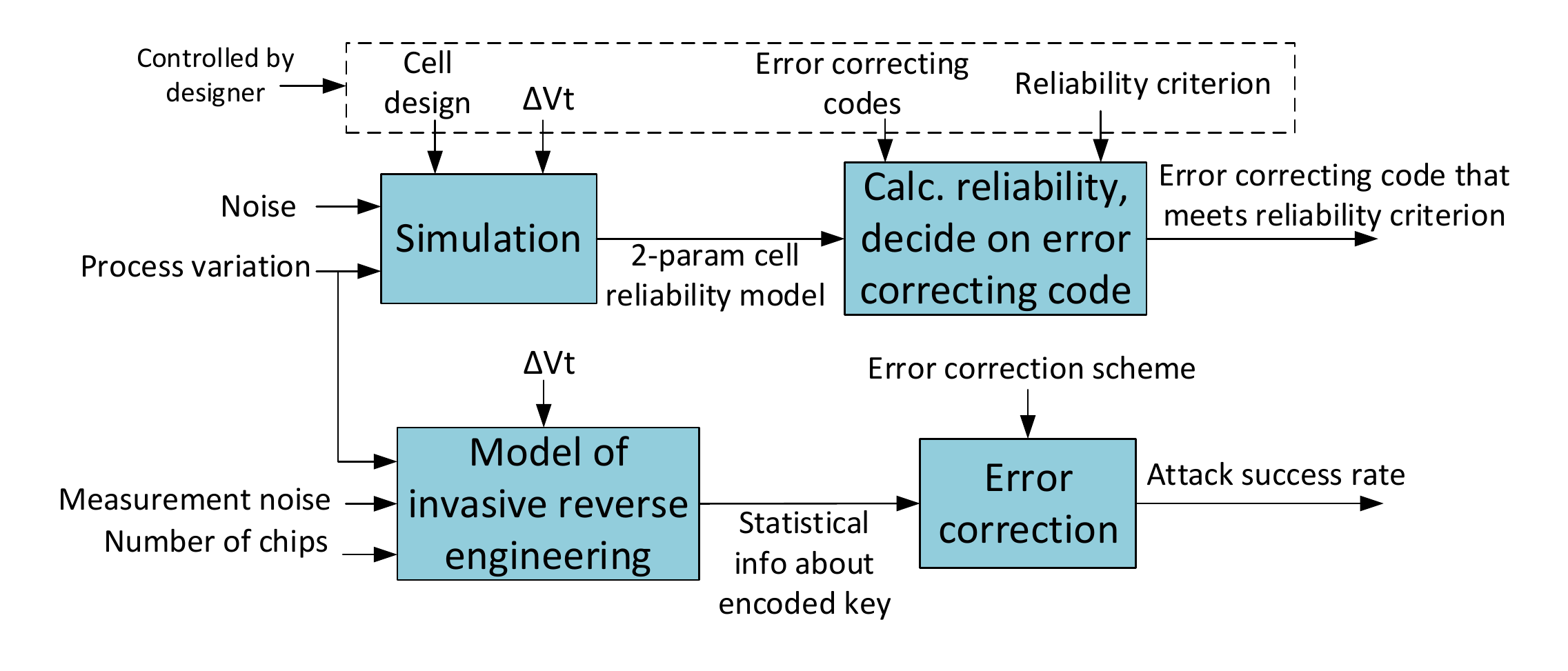}
	\caption{Overall CAD flow of the work}
	\label{fig:cad_flow}
\end{figure}

Figure~\ref{fig:cad_flow} shows our overall design flow. An engineer can use this framework to implement obfuscated keys that achieve desired tradeoffs of cost, reliability, and security. Each step of the flow is described in detail in the following sections. In the first step the designer chooses a cell type to use for storing the obfuscated constants, and chooses a threshold voltage offset to use for biasing the cells to generate codewords of the encoded key; from this, a cell reliability model is extracted. In the second step, the designer uses the cell reliability model and the chosen key reliability criteria to decide which error correcting code strengths are compatible with the circuit design. A candidate design then exists, and in the third step its security against invasive readout attack is quantified. Depending on whether the security level is deemed adequate, the design can be revised. Examples of revision can be to trade cost against security by decreasing threshold offset and increasing error correction, or trade reliability against security by using a weakened error correction. 

\section{Threshold-Based Key Storage Elements}
\label{sec:structure}

Threshold voltages of transistors are commonly chosen for power-performance tradeoffs, but recent works have shown that modifications to threshold voltages can also be used to determine the logical function of cells. The basic idea behind these works is to use multiple classes of transistors with different threshold voltages. The modification of threshold voltages can permanently make transistors on/off or can adjust their relative characteristics to cause the circuit to implement a specific function \cite{NirmalaThreshold, Erbagci2016threshold, collantes2016threshold, keshavarz2017design}. 

There exist a number of ways in which threshold voltages can determine digital values produced in a circuit. In another context, intrinsic variations in threshold voltages have been used to create device-tied identifiers~\cite{lofstrom-00,su-07} or secret values in PUFs. It has been previously shown that the power-up state of an SRAM cell depends on the intrinsic threshold voltage differences between transistors which are caused by process variation~\cite{SRAMpower}. In the same way that intrinsic differences in threshold voltages can randomly bias cells toward generating specific values, intentional differences in threshold voltages can bias cells toward specific values in a way that is common across chip instances. 

\begin{figure}
	\centering
	\includegraphics[width=0.7\columnwidth]{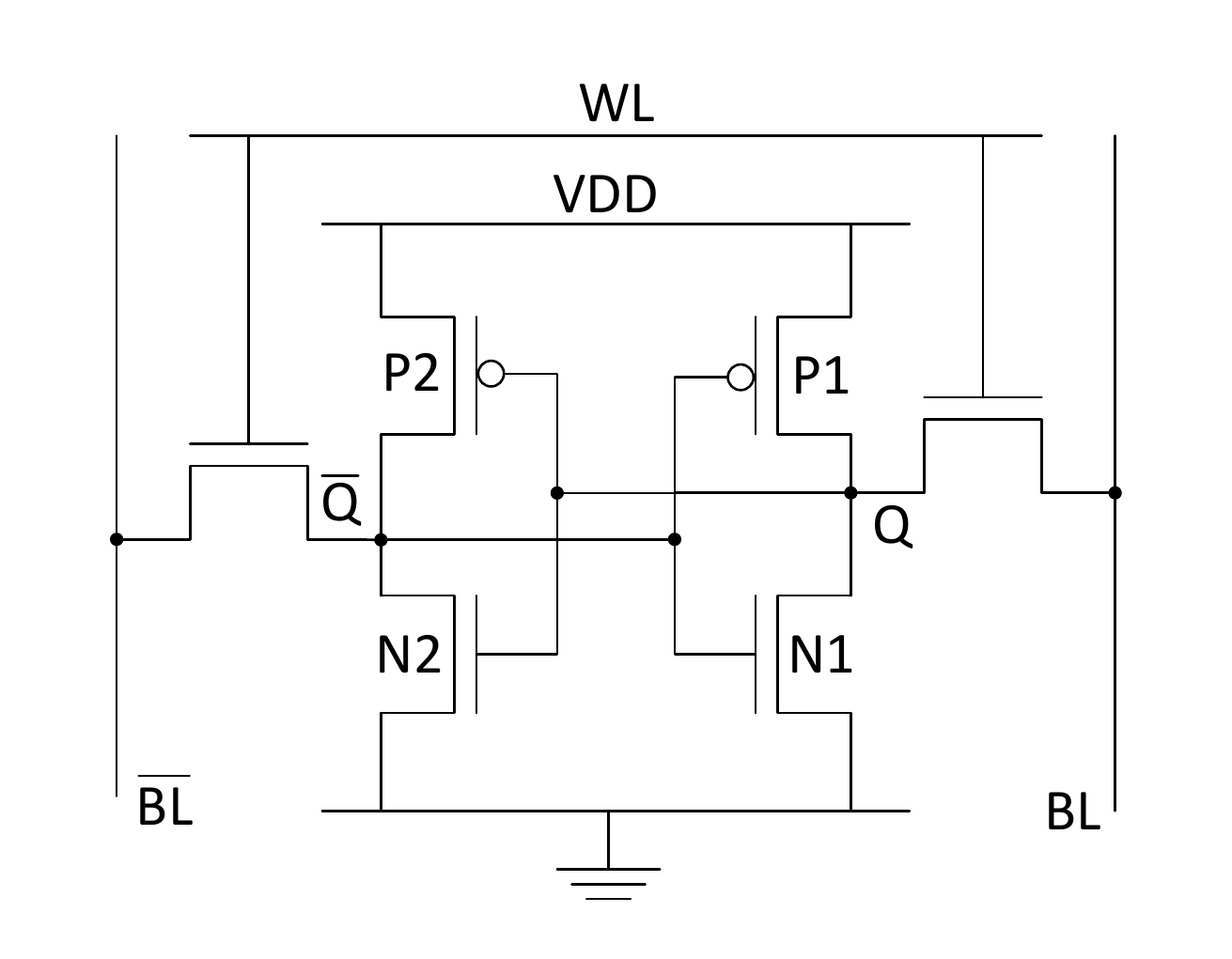}
	\caption{A simple 6T SRAM cell. The cell is biased toward the 1-state by increasing the magnitude of transistor P2, and biased toward the 0-state by increasing the magnitude of transistor P1.}
	\label{fig:SRAM}
\end{figure}

Consider a designer that wants to modify the 6-T SRAM cell of figure~\ref{fig:SRAM} so that it will generate a certain value each time it is powered up.
Without loss of generality, we assume the desired state is the 1 state ($Q=1,\overbar{Q}=0$), while noting that the 0 state works the same way due to the symmetry of the cell. For simplicity, we assume the designer wants to induce the desired state by changing the threshold voltage of only one transistor in the cell. Then the question arises regarding which transistor should be changed, and by how much should its threshold be changed. 

The designer can change the threshold voltage of one transistor in the cell in the following ways to bias the cell toward the 1 state: 
\begin{inparaenum}[\itshape 1)]
		\item increase the magnitude of threshold voltage on $N1$; 
		\item decrease the magnitude of threshold voltage on $N2$; 
		\item increase the magnitude of threshold voltage on $P2$; or
		\item decrease the magnitude of threshold voltage on $P1$.
\end{inparaenum} 
Regardless of which transistor threshold is modified, a larger magnitude change will make the cell more reliably biased, but will also give the attacker a better chance of correctly measuring the threshold difference invasively during reverse engineering. The designer therefore seeks to maximize the reliability that can be obtained for a given amount of threshold offset. 
\par
\begin{figure}
	\centering
	\includegraphics[width=0.8\columnwidth]{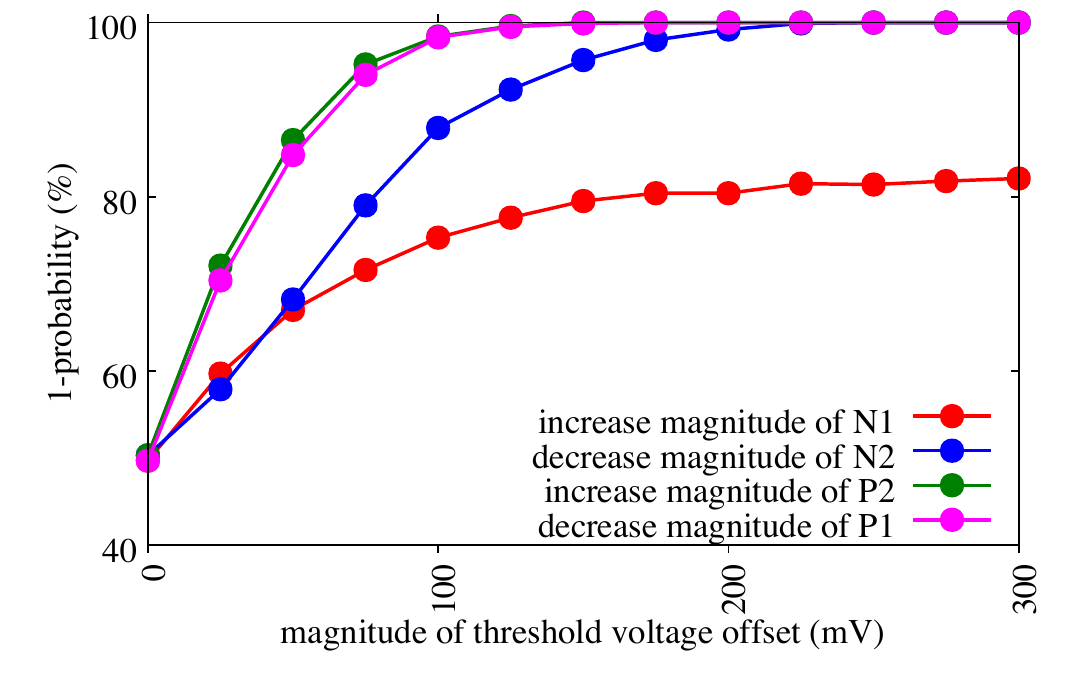}
	\caption{The 1-probability against different threshold voltage offsets for a PMOS and NMOS transistor}
	\label{fig:1-probability}
\end{figure}

To determine which transistor should be modified, we evaluate the 1-probability of the SRAM cell versus its threshold offset. 
For any threshold offset, the 1-probability shows the fraction of cells that are biased toward producing the desired 1 state after process variations are added. The evaluation is based on 1000 Monte Carlo simulation instances of an SRAM cell in HSPICE using 45nm CMOS Predictive Technology Model~\cite{PTM} (PTM) with nominal threshold voltage of 469mV for NMOS and -418mV for PMOS transistors. We consider standard deviation of threshold voltage distribution to be 30mV. The result of this comparison is shown in figure \ref{fig:1-probability}. It can clearly be seen that biasing the threshold of PMOS transistor results in a higher 1-probability. Therefore, we conclude that adding a threshold offset on the PMOS transistor is a more effective way to influence the value generated by the cell, compared to the same threshold offset on an NMOS transistor. Based on this analysis, the threshold-based cell programming that we use is to increase the magnitude of $P2$ to induce a 1 value in the cell. Because of the symmetric structure of an SRAM cell, a 0 is stored in the complementary way, by increasing the magnitude of $P1$. 



We have shown that it is more beneficial in terms of value stability to choose PMOS over NMOS transistors. Therefore, our technique requires two different kinds of PMOS devices with two different thresholds to choose from. One is nominal, and we assume in several places that the second is a threshold of our choosing. Multi-threshold processes are common, but different fabrication processes will typically offer fixed choices for thresholds. There are no technical barriers to having the second threshold be arbitrary, and for the sake of exploring the achievable limits of obfuscation, we will assume fabrication cooperation that allows us to freely choose threshold. For a slightly more granular approach, a designer can choose among a discrete number of thresholds that are available in existing commercial processes. 

\section{Reliability of Threshold-based Keys}
\label{sec:key_reliability}
In cryptography, even a single key bit upset may cause discernible consequences, and threshold-programmed values are inherently unreliable due to noise and process variation. For this reason, our scheme uses error correction in addition to the threshold-programmed cells. The parameters of error correction and the cell threshold offsets must be chosen together to ensure that the key meets a reliability criterion; a larger threshold offset improves cell reliability and allows weaker error correction to suffice, while a smaller threshold offset will require a correspondingly stronger error correction. Due to variations across chips, the chips will not all have the same key failure rates. Any reliability criterion must therefore specify both a key failure rate, and a fraction of chips that must have key failure rates below that number. The reliability criterion that we use is that at least 99\% of chips must have a key failure rate of less than $10^{-6}$. The error correcting code selected for any threshold offset must cause this criterion to be satisfied. 

Because key failures are such infrequent events, it is not possible to check whether a design meets the given reliability criterion using random simulation alone, so we rely on a careful combination of simulation, modeling, and statistics. The scheme we use to check reliability is a two step process. The distribution of cell error probabilities is first captured in a two-parameter abstracted model. The model of cell error probabilities is then used within a procedure that calculates the distribution across chips of the key failure rate for different error correction schemes.

\subsection{Distribution of Error Probabilities across Cells}

Each cell is biased to produce a single 0 or 1 bit of a codeword as chosen by the designer, but due to noise and variability it may not produce this desired value in a given trial. In fact, due to process variations, some cells may almost never produce the desired value, while other cells will produce it sometimes or almost always. Circuit simulation is used to learn the distribution of cell error probabilities for a given threshold offset.

\par
Our baseline data for cell reliability is generated using HSPICE simulation of SRAM cells in 45nm Predictive Technology Model (PTM). We created 512 SRAM cell instances with variation on transistor threshold voltages according to PTM, and evaluated each cell in the presence of transient noise 300 times. Noise is captured in the simulations of each instance by doing a single-sample Monte Carlo transient noise analysis with the $.TRANNOISE$ command. From these simulations, a set of empirical cell error probabilities is obtained.

\begin{figure}
	\centering
	\includegraphics[width=0.8\columnwidth]{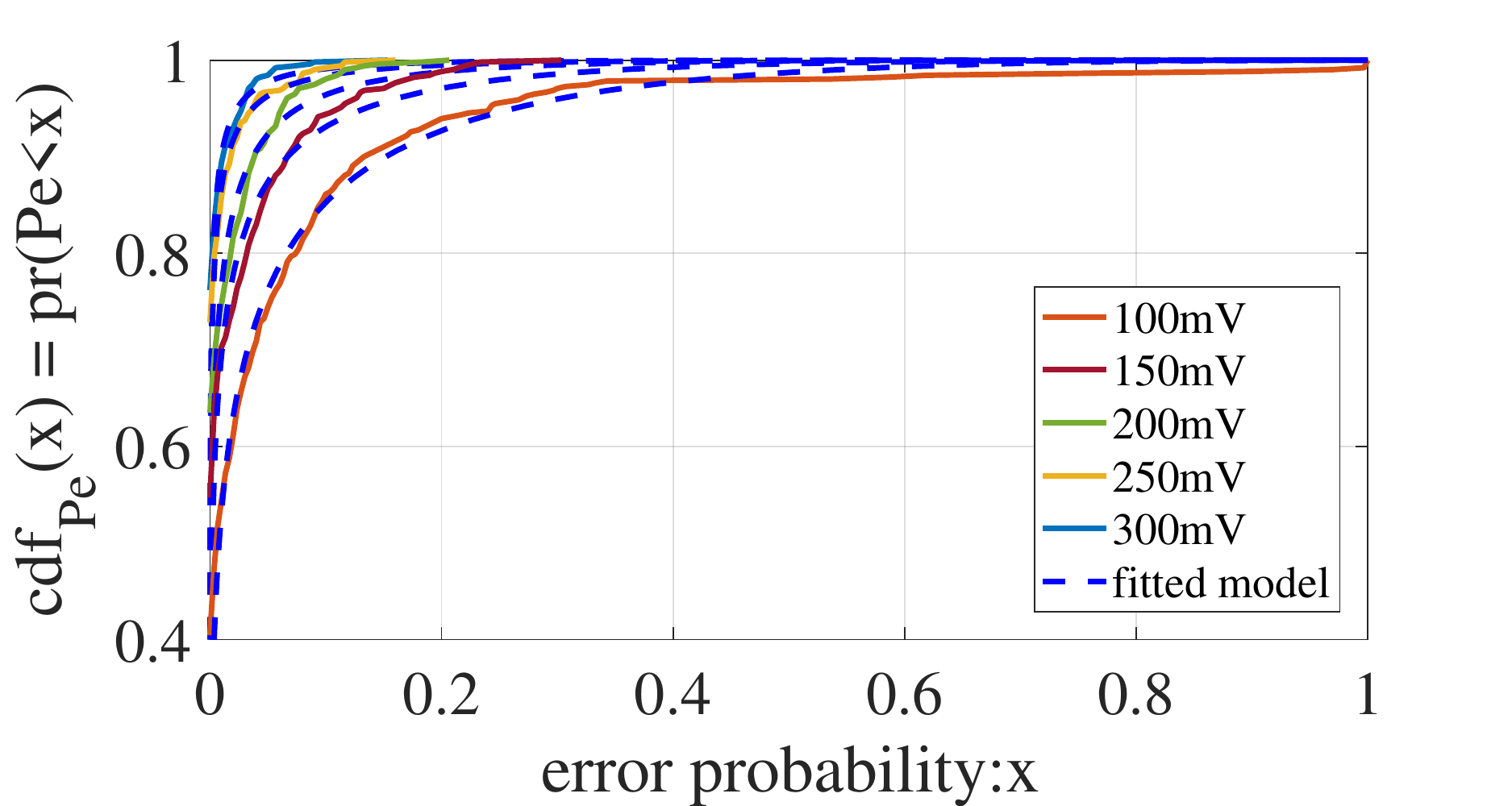}
	\caption{Cumulative distribution function of error probabilities from the simulation data and their relative fitted curves for different magnitudes of voltage offsets}
	\label{fig:fitting}
\end{figure}

Noting that an SRAM cell with an intentionally offset threshold voltage is similar to a biased PUF, we adopt a modeling approach from PUFs to compute an expression that describes the distribution of cell error probabilities. 
The heterogeneous error rate model we use is proposed for PUFs by Roel Maes~\cite{Maes_2013}. The model assumes two sources of variation in a cell: The process variable ($M$) that models the persistent impact of bias and process variations, and the noise variable ($N_i$) that accounts for the cumulative effect of all noise sources during evaluation. Both variables are normally distributed. The process variable has an unknown mean and variance, while the noise variable is modeled as having 0-mean and an unknown variance. These three unknowns reduce to two unknown parameters $\lambda_1$ and $\lambda_2$ in the model (Eq.~\ref{eq:errorProb}); $\phi(x)$ and $\phi^{-1}(x)$ represent the cumulative distribution function of standard normal distribution x and its inverse, respectively. Parameters $\lambda_1$ and $\lambda_2$ are chosen by fitting Eq.~\ref{eq:errorProb} to the empirical CDF of cell error probability from circuit simulation using Levenberg-Marquardt algorithm. Figure~\ref{fig:fitting} shows the fitting of the model to simulation data for various threshold offsets. Having an expression for the distribution of cell error probabilities ($Pe$) allows us to sample from this distribution in order to obtain representative cell error probabilities. 

\begin{equation}
\label{eq:errorProb}
\textbf{cdf}_{Pe}(x) = \phi(\lambda_1 \phi^{-1}(x) + \lambda_2)
\end{equation}

\subsection{Distribution of Key Failures Across Chips} \label{ss:keyFail}
For any threshold voltage offset, using the known distribution of cell error probabilities, we can compute the distribution of key failure rates that will be achieved using different error correcting codes, and can check which codes satisfy our reliability criterion. We focus on BCH codes, which is a class of codes with different block sizes and numbers of correctable errors in each block. We denote a certain BCH code as $BCH[n, m, t]$; where $n$ is the block size, $m$ is the number of useful information bits per block after error correction, and $t$ is the number of correctable errors in each block. In our setting, $n$ is the number of SRAM bits used to store a portion of the encoded key, $m$ is the number of key bits generated from decoding the $n$ bits, and $t$ is the maximum number of SRAM bit errors that can be tolerated. If an error correcting code is able to correct $t$ bits, the block fails if more than $t$ bits are erroneous.

\par
The number of blocks required to generate a key using a given BCH code will depend on the desired key size ($k$) and the number of useful information bits from each block in that BCH code ($m$). The number of blocks needed for key generation is therefore $\lceil{\frac{k}{m}} \rceil$. The key generation fails if at least one code block that contributes to the key fails. If $P_{Fblock,i}$ is the probability of failure in block $i$, then the key failure probability is given by Eq.~\ref{eq:keyfail}.

\begin{equation}
\label{eq:keyfail}
P_{Fkey} = 1-\prod_{i=1}^{\left \lceil{\frac{k}{m}}\right \rceil } (1-P_{Fblock,i})
\end{equation}



\par
For each block of $BCH[n, m, t]$ code, the probability of producing an erroneous result is the probability that the number of errors in that block exceeds $t$. With a heterogeneous error rate model of cells, each block in a chip will have a failure rate that depends on the unique error rates of its cells. 
Hence, we cannot use binomial distribution to find failure rate of each block and instead, we use a more general case of binomial distribution, called ''Poisson-binomial distribution''. The distribution is a discrete probability distribution to calculate summation of Bernoulli trials that are not necessarily identically distributed. Given a set of $n$ non-uniform cell error rates $P_e^n=(p_{e,1}, p_{e,2},...,p_{e,n})$ in a block, the probability of having less than $t$ errors is calculated using cumulative distribution function of Poisson-binomial distribution $F_{PB}(t;P_e^n)$ as shown by Maes~\cite{Maes_2013} and given by Eq.~\ref{eq:poissonBinom}; this describes the probability of correctly decoding the block. Therefore, the failure rate the same block is given by Eq.~\ref{eq:pblock}

\begin{equation}
\label{eq:poissonBinom}
	\begin{multlined}
		F_{PB}(t;P_e^n) = \frac{t+1}{n+1}+\frac{1}{n+1}{\sum_{m=1}^n\frac{(1-e^{\frac{-j2\pi m(t+1)}{n+1}})}{(1-e^{\frac{-j2\pi m}{n+1}})}} \\ .\prod_{k=1}^{n}(p_{e,k}e^{\frac{j2\pi m}{n+1}}+(1-p_{e,k}))
	\end{multlined}
\end{equation}

\begin{equation}
\label{eq:pblock}
P_{Fblock} = 1-F_{PB}(t;P_e^n)
\end{equation}

We now describe the steps to use the equations given above for evaluating key reliability with a given BCH code and given threshold offset. First, we sample cell error probabilities from the fitted $\textbf{cdf}_{Pe}$ (Eq.~\ref{eq:errorProb}) using inverse transform sampling to obtain a set of $n$ representative cell error probabilities ($P_e^n$); because the error probabilities are fitted to simulation results, this accounts for circuit-level reliability. We repeat the sampling for the number of required blocks, and then for each one calculate the block failure rate ($P_{Fblock}$) using Eq.~\ref{eq:poissonBinom} and Eq.~\ref{eq:pblock}, and use the block failure rates to compute the key failure rate using Eq.~\ref{eq:keyfail}. This calculated key failure rate is for one chip instance with a specific combination of threshold offset and BCH code. Repeating the whole calculation multiple times produces the distribution of key failure rates, and we use that distribution to evaluate whether the combination of threshold offset and BCH code satisfy our reliability criterion of at least 99\% of chips having key failure rates of less than $10^{-6}$.



\begin{figure}
	\centering
	\subfloat[Key failure rates of different BCH error correcting codes, for threshold offset magnitude of 200mV]{
	\includegraphics[width=0.95\columnwidth]{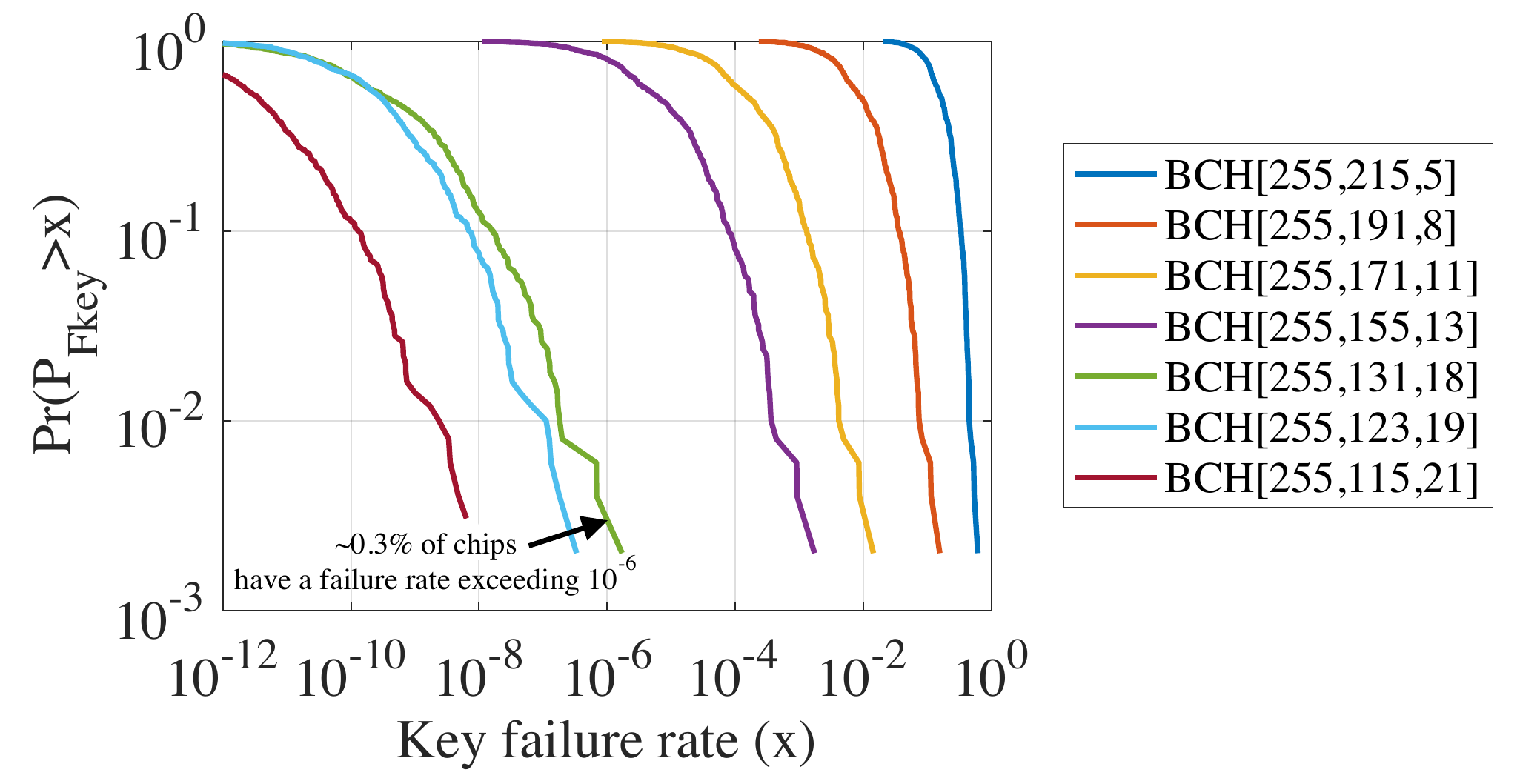}
	\label{fig:KeyFail}
	}
	\newline
	\centering
	\subfloat[Key failure rates of all (threshold offset,BCH code) pairs that meet the reliability criterion]{
	\includegraphics[width=0.95\columnwidth]{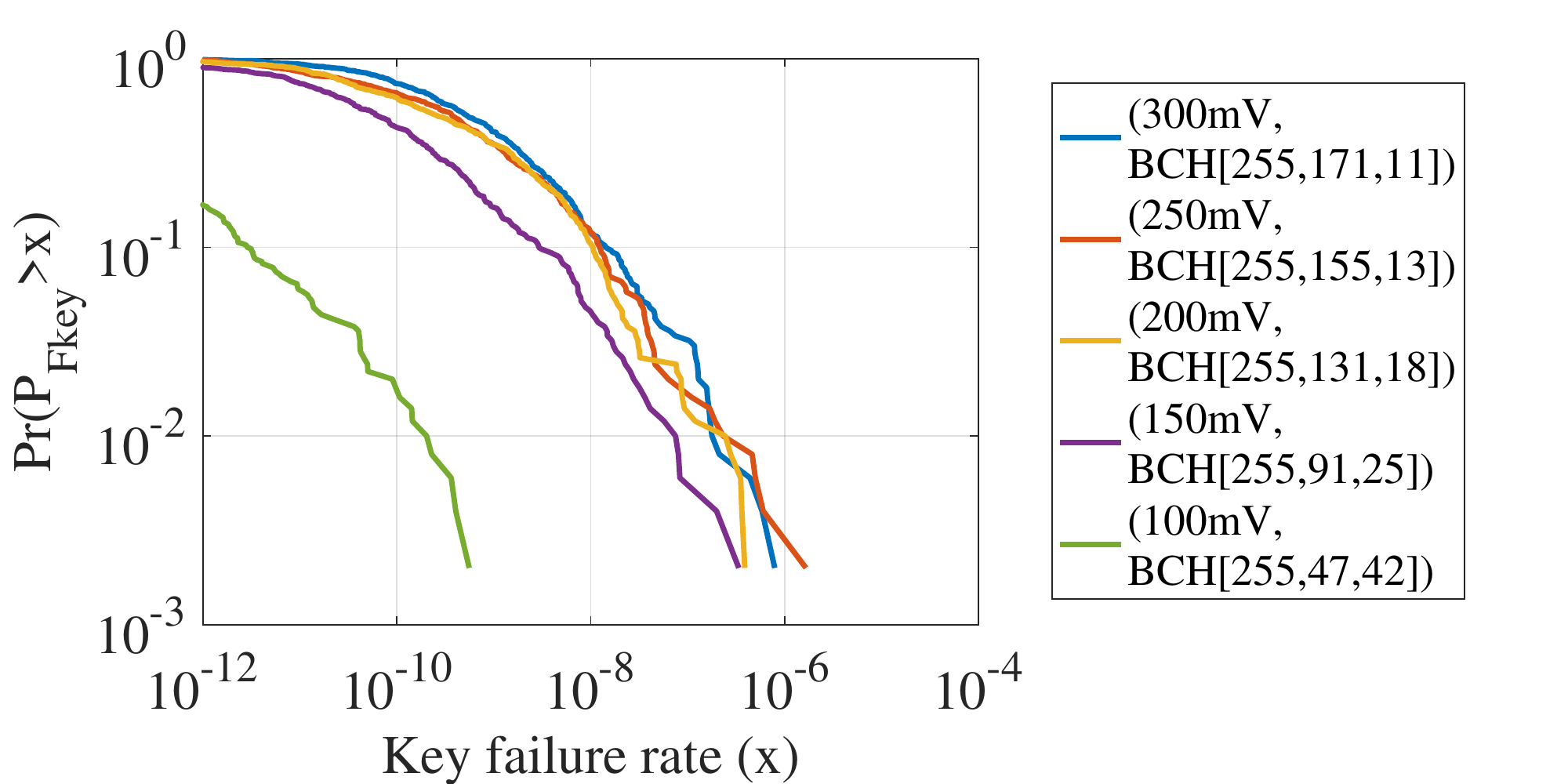}
	\label{fig:OffsetBCH}
	}
	\raggedleft
	\caption{Key failure rates of different design options}
	\label{fig:keyFailure}
\end{figure}

Among all the BCH codes that will satisfy our reliability criterion for a given threshold offset, we use only the lowest cost BCH code, which is the one that corrects the fewest errors among all sufficiently reliable codes. A designer can choose from different combinations of threshold voltage offsets and error correcting codes to reach the desired reliability for the key. Figure \ref{fig:KeyFail} shows the key read failure rate for threshold offset of magnitude 200mV, evaluated for different BCH codes. As can be seen, the least expensive code that meets our reliability criterion is $BCH[255,131,18]$.

Figure \ref{fig:OffsetBCH} shows, for each threshold offset, the distribution of key failure rates that occurs when the minimal BCH code meeting the reliability criterion is used. Table \ref{tab:BCH_codes} shows the area of each of these combinations in order to generate a 128-bit key that satisfies the reliability requirement of at least 99\% of chips having key failure rate of less than $10^{-6}$. For our area overhead evaluations we used SRAM cell area of 0.345$\mu m^2$ as reported in \cite{SRAMSize} and synthesized the BCH decoders using NanGate 45nm Open Cell Library \cite{NanGate}. The cost of each option is provided in terms of area in $\mu m^2$ units. Given that equivalent reliability can be obtained by these different combinations of threshold offset and BCH code, one must consider the implications of choosing among the equivalent-reliability design alternatives. As we will show in the next section, each of these approaches comes with some tradeoff of cost and security. Using a higher threshold voltage offset makes reverse engineering easier, but using a stronger error correcting code comes with more expense in terms of area and power consumption.


\begin{table*}[]
	\centering
	\caption{Evaluation of equivalent-reliability designs. Each pairing of threshold offset and BCH code are chosen such that the BCH code is the lowest cost code that will satisfy the reliability criterion for that threshold offset.}
	\label{tab:BCH_codes}
\begin{tabular}{|l|l|l|l|l|l|}
	 \hline
	$\Delta_{vt} (mV)$ & 100     &150	& 200        & 250         & 300         \\ \hline
	\multicolumn{1}{|l|}{\begin{tabular}[c]{@{}l@{}}BCH code parameters (n, m, t)\end{tabular}} & (255,47,42)  & (255,91,25) & (255,131,18) & (255,155,13) & (255,171,11) \\ \hline

	\multicolumn{6}{l}{}  \\ \hline
	\multicolumn{1}{|l|}{Number of cells to store encoded key}          & 765      & 510	& 255          & 255           &  255      \\ \hline
	\multicolumn{1}{|l|}{Cells area overhead ($\mu m^2$)   }      & 264      & 176 	&	 88       & 88           &  88      \\ \hline 
	\multicolumn{1}{|l|}{BCH decoder area ($\mu m^2$) }         & 61403      & 40723	&	 31428          & 24835           &  21602      \\ \hline \hline	
	\multicolumn{1}{|l|}{Total area (SRAM cells + BCH decoder ($\mu m^2$))}               & 61667 &40899	         & 31516    &24923       &  21690    \\ \hline
	\multicolumn{6}{l}{}  \\ \hline
	\multicolumn{1}{|l|}{Attacker success for a single chip ($RS_{key}$)}                        & 8.99e-36  	&1.45e-28    & 5.26e-13     & 6.90e-11      & 7.66e-08   \\ \hline
\end{tabular}
\end{table*}

\section{Resistance Against Invasive Readout}
\label{sec:reverse_eng}
If the designer uses the proposed technique to store a key, the first question that comes to mind is how resistant this key is to reverse engineering attacks. As explained in the previous sections, our approach benefits from the use of error correcting codes to correct the impact of noise and manufacturing issues on key values. The strength of this code is chosen in accordance with the threshold offset ($\Delta_{Vt}$); a smaller threshold offset will require stronger error correction to reach its desired reliability. Choosing a small threshold offset makes it harder for a reverse engineer to distinguish between the different measured threshold voltage values, but the stronger error correction can also help the attacker to correct errors in his own invasive measurements. This makes it difficult for a designer to increase security without compromising reliability, and leads to a space of trade-offs between reliability, security and cost that must be considered during design. In this section, we will evaluate the resistance of each design option against reverse engineering. 




\subsection{Attacker Model}\label{ss:attacker_model}

We conservatively assume that an attacker knows everything about the encoded secret key except for the key value that the designer has encoded. The attacker knows which cells store the encoded values, and knows that the secret key bits are encoded into the cells by increasing the magnitude of threshold voltage on either transistor $P1$ or $P2$ to encode a 0 or 1 bit. The attacker also knows the parameters of the BCH error correction that is used.
 \par
Using this knowledge to reverse engineer the encoded values, the attacker has to somehow guess enough bits correctly that applying the error correction to his guess will produce the key. For example, if the designer added a BCH error correcting block capable of correcting $t$ errors, the attacker's guess of the encoded key must be within $t$ bits of the value that the designer intended to store. The attacker learns about encoded key bits by invasively measuring the threshold voltages of $P1$ and $P2$ to guess whether the cell stores a 0 or 1 value. 
\par
Since threshold voltage cannot be learned through conventional methods such as delayering and imaging, most works on multi-threshold obfuscation regard the threshold voltage as being perfectly secure. However, there are still methods such as spreading resistance profiling (SRP) \cite{mazur1966SRP}, scanning capacitance microscopy (SCM) \cite{kopanski1996SCM}, scanning spreading resistance microscopy (SSRM) \cite{SSRM} and Kelvin probe force microscopy (KPFM) \cite{koren2009nonuniform} to measure the concentration of dopant atoms in the channel and hence reveal the threshold voltage. However, these methods still have low read accuracy and high overhead. We evaluate the key stealthiness even for high threshold read accuracies that may not be feasible yet. 

\par
Regardless of the technique used to invasively measure transistor threshold voltages, there will be some imperfection to the measurements.
 Measuring the threshold of transistors and their relative values can be a difficult task since the measurement precision of threshold voltages may not be perfect, and even the task of preparing the chip for measurement can be difficult. There are two sources of inaccuracy that limit the attacker's success in reverse engineering the obfuscated key:
\begin{enumerate}
\item \textbf{Manufacturing Variations}: Process variations cause the threshold voltages of manufactured transistors to differ from the nominal values intended by the designer. The effect of process variation on threshold voltage of each transistor has a distribution of $\mathcal{N}(0, \sigma^2_{var})$. This is the same process variation model used in circuit simulation in Section~\ref{sec:structure}. 
\item \textbf{Measurement Error}: Regardless of the type of measurements performed by the attacker to read out the threshold voltages, some inaccuracy is inevitable. Measurement error causes the reverse engineer to measure a threshold voltage that differs slightly from the true threshold of the transistor. We model measurement error as $\mathcal{N}(0, \sigma^2_{err})$.
\end{enumerate} 


\par

Consider the attacker's view of a cell that is designed to store a 1. The magnitude of threshold voltages of $P2$ and $P1$ are $\mathcal{N}(vt+\Delta_{vt}, \sigma^2_{var})$ and $\mathcal{N}(vt, \sigma^2_{var})$ respectively, because an increased threshold on $P2$ is the mechanism used to create a 1-value. The threshold voltages of $P1$ and $P2$ as read by the attacker with measurement error are $\mathcal{N}(vt+\Delta_{vt}, \sigma^2_{var} + \sigma^2_{err})$ and $\mathcal{N}(vt, \sigma^2_{var} + \sigma^2_{err})$ respectively. The attacker should guess that the cell stores a 1 value if he measures a higher threshold voltage on $P2$. The difference between the measured threshold voltages of $P2$ and $P1$ is $\mathcal{N}(\Delta_{vt}, 2\sigma^2_{var} + 2\sigma^2_{err})$, and when this difference is positive, the attacker guesses a value for the cell that is the same as what the designer intended for the cell. The probability ($P_{re}$) that the attacker will infer the wrong value for the cell is then the cumulative distribution function of $\mathcal{N}(\Delta_{vt}, 2\sigma^2_{var} + 2\sigma^2_{err})$ evaluated at point $x=0$ (Eq.~\ref{eq:perr}).


 \begin{equation}
 \centering
 \label{eq:perr}
 P_{re} = \textbf{cdf}_{\mathcal{N}(\Delta_{vt}, 2\sigma^2_{var} + 2\sigma^2_{err})}(x=0)
 \end{equation}

\begin{figure}
	\centering
	\includegraphics[width=0.9\columnwidth]{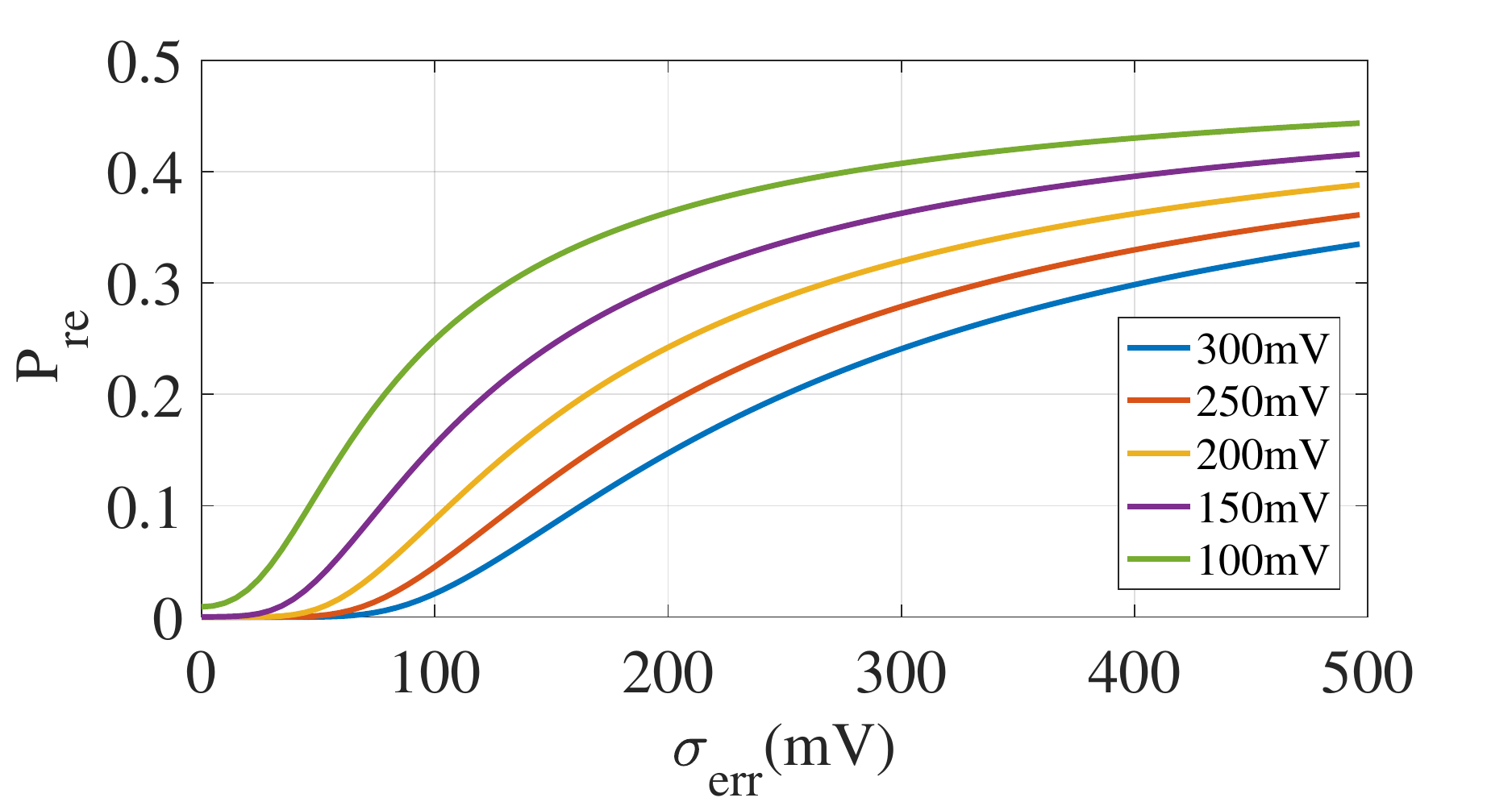}
	\caption{For different values of $\Delta_{vt}$, plot shows the probability ($P_{re}$) that an attacker reads a value for a cell that differs from the value programmed by the designer, as a function of the attacker's measurement error ($\sigma_{err}$).}
	\label{fig:perr_vs_vth}
\end{figure}

Figure \ref{fig:perr_vs_vth} shows the probability, for different values of $\Delta_{vt}$ and $\sigma_{err}$, of an attacker inferring a value that disagrees with the value intended by the designer. As would be expected, this probability of misreading a cell is higher when the threshold offset ($\Delta_{vt}$) is small or the standard deviation of measurement error ($\sigma_{err}$) is large.  

\subsection{Attacker's Success Rate for Key Readout} \label{ss:readKeyFail}


To correctly guess the key, the attacker has to guess the encoded key bits with a number of errors that is within the error correcting capacity of the BCH code. Having the probability of cell read error ($P_{re}$) from Eq.~\ref{eq:perr}, the number of errors in a block is binomially distributed, and the probability of the attacker successfully reading out a single block of a $BCH[n,m,t]$ error correcting code is given by Eq.~\ref{eq:readblocksuccess}.

 \begin{equation}
\centering
\label{eq:readblocksuccess}
P_{RSblock} = \sum_{i=0}^t {{n}\choose{i}} (P_{re})^i(1-P_{re})^{n-i}
\end{equation}

Given that multiple error correction blocks may be required to generate the entire key, the attacker will only succeed in reading out the key when all blocks are read correctly. The probability of the attacker reading out the key successfully is denoted $P_{RSkey}$ and calculated as shown in Eq.~\ref{eq:keyreadsucc}. Table \ref{tab:BCH_codes} reports the attacker success rate ($P_{RSkey}$) for different threshold offset magnitudes when $\sigma_{err} = 200mV$.

\begin{equation}
\label{eq:keyreadsucc}
P_{RSkey} = \prod_{i=1}^{\left \lceil{\frac{k}{m}}\right \rceil } P_{RSblock,i}
\end{equation}



\subsection{Cost of Readout by Attacking Multiple Chips}


When the same key is encoded in multiple chips, an attacker can choose to attack multiple chips in order to improve accuracy by averaging out deviations in measurement error and process variations. In this case, the attacker sees the differences between the transistor threshold voltages in a cell as $\mathcal{N}(\Delta_{vt}, \frac{2\sigma^2_{var} + 2\sigma^2_{err}}{C})$, where $C$ is the number of chips measured. Changing the normal distribution of Eq.~\ref{eq:perr} to account for this reduced variance leads to a reduction in $P_{re}$ which benefits the attacker. Note that taking measurements from additional chips is preferable over taking multiple measurements of the same chip, which only reduces measurement noise but not process variations. Depending on the costs of preparing a chip for measurement, there could be advantages to re-measuring a single chip, but we do not consider that here.


\par
\begin{figure}
	\centering
	\includegraphics[width=0.9\columnwidth]{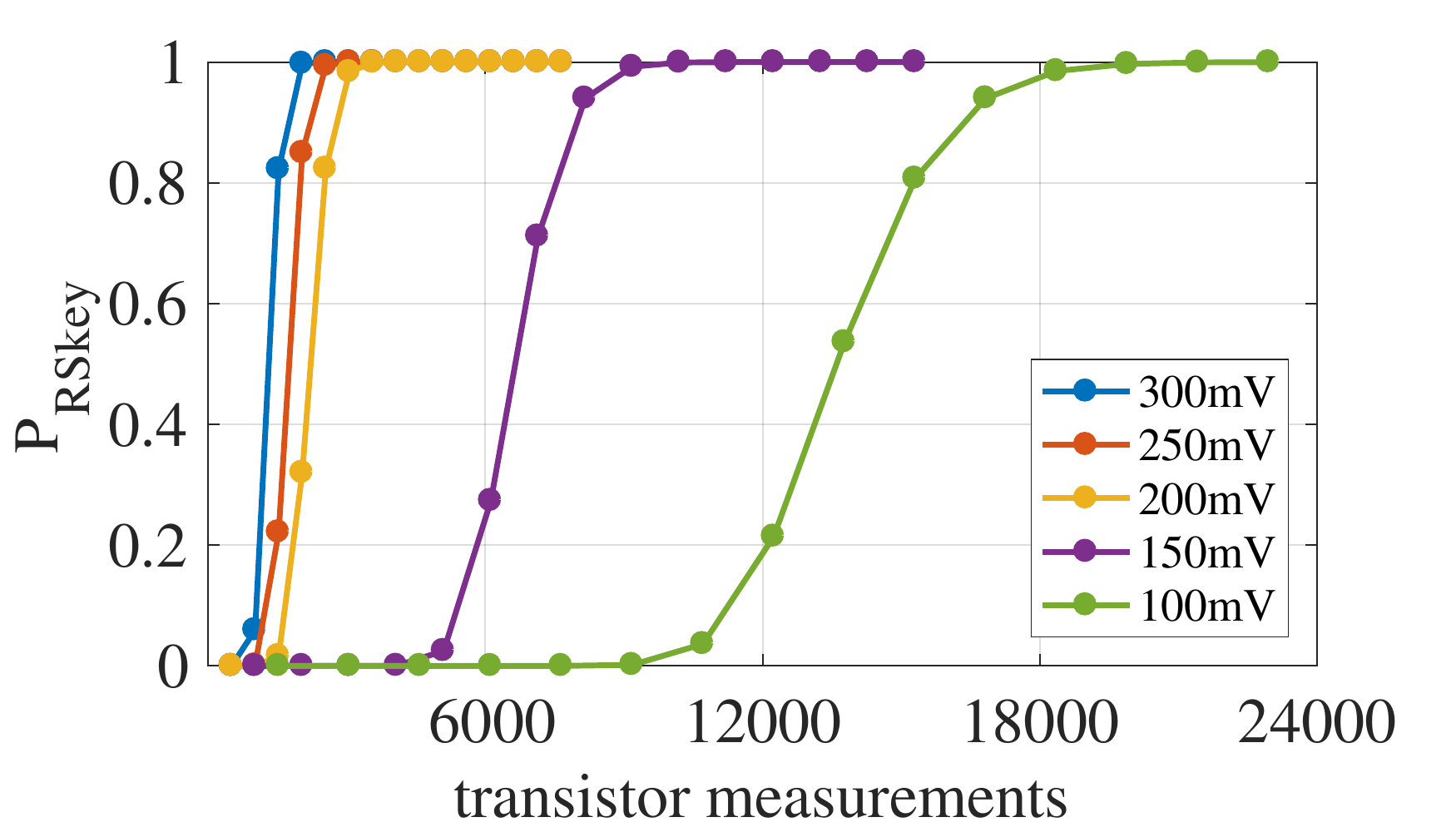}	
	\caption{Effect of multiple chip measurements on reverse engineering success rate for different threshold offsets with $\sigma_{err}$ = 200mV }
	\label{fig:ImproveVsMeasurements}
\end{figure}

 Figure \ref{fig:ImproveVsMeasurements} shows the relation of reverse engineering success rate with the number of individual chips used for measurements for each threshold offset. As an example, one can observe that when the threshold offset ($\Delta_{Vt}$) is 100mV, the attacker has to measure about 13770 transistors to have more than a 53\% chance of extracting the key. This requires measuring two transistors from all 6885 cells that store the encoded key on 9 instances of the chip.'
However, it should be noted that although having more chips increase the attacker's success rate, it also comes with extra cost of measuring multiple threshold values.

As mentioned before, parameters of a BCH code are denoted as $[n,m,t]$ where $n$ is the block size, $m$ is the size of useful data after error correction and $t$ is the size of correctable errors in a block. 
For a key of size $k$ that uses BCH blocks of size $[n, m, t]$, a total of $\lceil \frac{k}{n} \rceil$ blocks are used. Therefore, there are $m * \lceil \frac{k}{n} \rceil$ input bits for BCH blocks that are provided by threshold-biased SRAM cells. The reverse engineer needs to measure the threshold of two PMOS transistors for each cell, making a total number of $2m * \lceil \frac{k}{n} \rceil$ transistor threshold measurements per chip in order to extract the key in this setting. 
\par
If the reverse engineer tries to increase the key read reliability by measuring the cell values from $C$ chips, it will increase the number of transistor threshold measurements to a total of $C * 2m * \lceil \frac{k}{n} \rceil$. In this way, using a smaller value of $\Delta_{vt}$ combined with stronger error correction has two advantages. By storing information more diffusely, it requires more measurements to be made on each chip, and requires more chips to be attacked before the key can be guessed.


\section{Design Tradeoffs}

Having shown analysis of reliability and security for different design scenarios, we now discuss how a designer can maximize her advantage over the attacker for effective security tradeoffs. 
While most changes will impact both reliability and security, some will represent more effective tradeoffs for the designer to consider.

\subsection{Loosening Reliability Constraints}
Error correcting code choice is constrained by our reliability criterion which specifies a maximum key failure rate for chips in the first percentile of reliability. In other words, we've specified that 99\% of chips must satisfy some reliability bound. If we allow weaker error correction to be used, then the failure rate of chips in the first percentile of reliability will increase. Yet, at the same time, the attacker's success rate for extracting the key will decrease. 

\begin{figure}
	\centering
	\includegraphics[width=0.9\columnwidth]{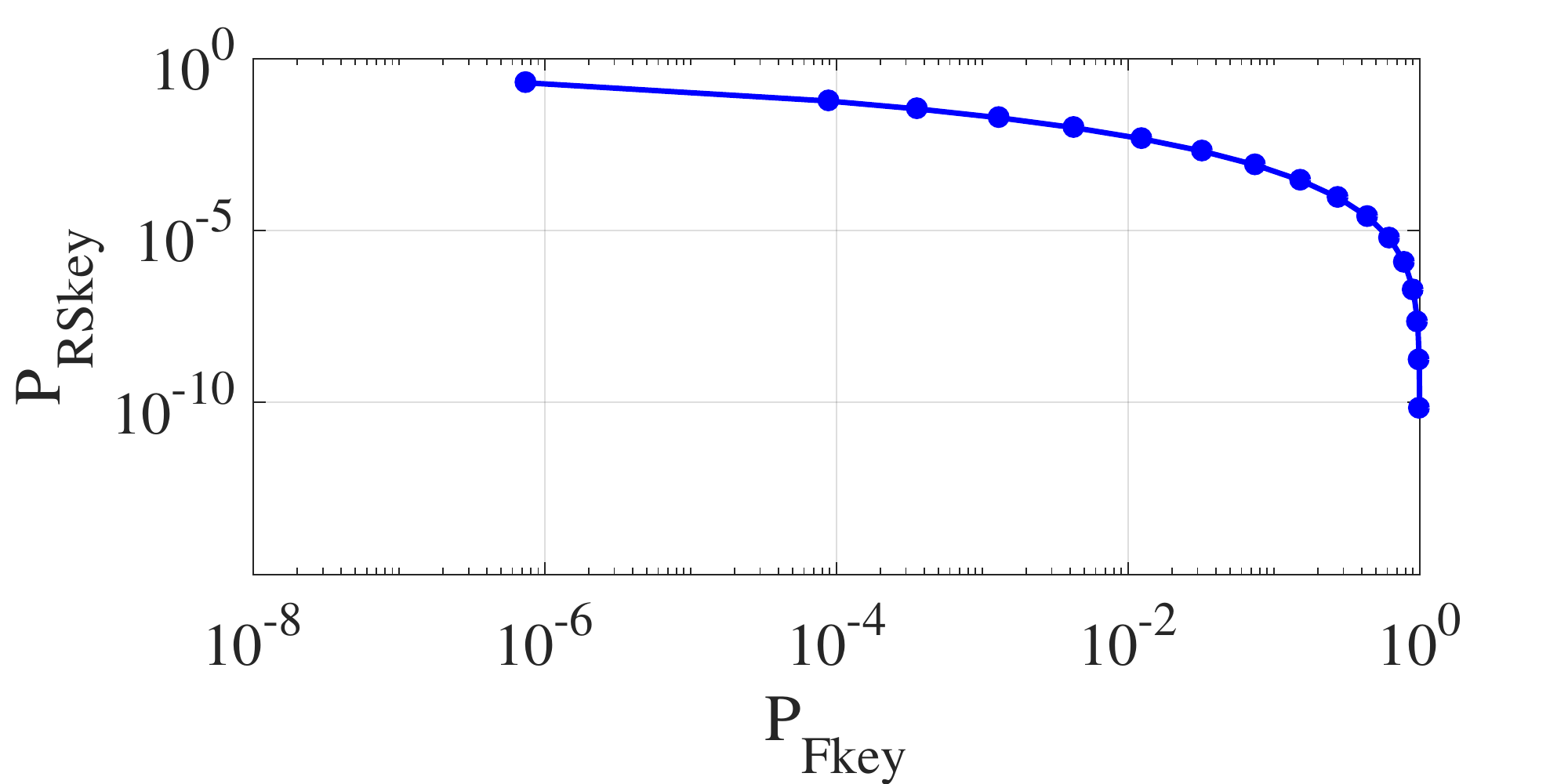}	
	\caption{Tradeoff between attacker's key read success rate ($P_{RSkey}$) and key failure rate ($P_{Fkey}$) that can be achieved by using different error correcting codes.}
	\label{fig:KFvsRS}
\end{figure}

To compare the key reliability of the design to the key read success rate of an attacker, Figure \ref{fig:KFvsRS} shows the security versus reliability tradeoff offered by different error correcting codes. This plot is analyzes a scenario with a threshold offset ($\Delta_{vt}$) of 200mV, and a low measurement error ($\sigma_{err}$) of 100mV. The leftmost point shows the attackers high success rate if the BCH code used is strong enough to ensure that 99\% of chips have an error rate less than 1E-6, as was used before. If different BCH codes are used, the plot shows how the failure rate of first-percentile chips increases, and the attacker success rate decreases, with increasingly weaker BCH codes. This curve represent a set of tradeoffs that a designer can make. Allowing a higher failure rate in key generation may be desirable in some scenarios if higher level error correction mechanisms occur. Note that this particular scenario is one in which the attacker is already able to make highly precise measurements, and that the achievable tradeoffs can be even better in other cases.


\subsection{Majority Voting}
Majority voting using multiple values obtained from each cell provides a way for the designer to mitigate the effects of on-chip noise. This an interesting tradeoff for the designer because on-chip noise, which is detrimental to key reliability, does not present any difficulty to the attacker since his read-out is not based on observing digital values from a functional chip. Therefore, majority voting is an attractive way to improve reliability of cell values and allow a weaker BCH code to be used, which has the effect of making the attacker's task more difficult without compromising key reliability. In other words, the designer can strategically replace some amount of algorithmic error correction that helps the attacker, with an amount of circuit-level error correction that does not help the attacker.

\section{Conclusion}
\label{sec:conclusion}
This work presents a methodology for storing obfuscated master keys with quantifiable security against an attacker that knows everything about the design except for the values of the secret key bits. The underlying technique is to combine threshold-based secrets with error correcting codes to allow secrets to be stored diffusely, which gives the designer an advantage over attackers that try to read out the secrets with some amount of imprecision. The proposed methodology enables designers to achieve different tradeoffs of area cost, key reliability, and security against invasive readout. Future work building on these ideas can consider even more diffuse ways to store keys, can consider technologies other than threshold voltages in SRAM, and can consider how to impart biases on cells as a post-manufacturing step so that the technique can be used for device-tied keys in addition to secure storage of master keys.
\par
\textbf{Acknowledgement}: This work has been supported by a grant from the National Science Foundation (NSF) under award CNS-1563829 and by University of Massachusetts, Amherst.

\thispagestyle{empty}


\bibliographystyle{acm}
\bibliography{refs}

\end{document}